 \documentstyle[11pt,aaspp4]{article}

\begin{document}

\title{THE METALLICITY AND DUST CONTENT OF HVC 287.5+22.5+240: EVIDENCE FOR A
MAGELLANIC CLOUDS ORIGIN\footnote{Based on
observations with the NASA/ESA Hubble Space Telescope obtained at the Space
Telescope Science Institute, which is  operated by Association of Universities
for Research in Astronomy, Inc., under NASA contract NAS5-26555.}}
\author{Limin Lu$^{2,3}$, Blair D. Savage$^4$, Kenneth R. Sembach$^5$, 
Bart P. Wakker$^4$,}
\author{Wallace L.W. Sargent$^1$, and Tom A. Oosterloo$^6$}
\altaffiltext{2}{California Institute of Technology, 105-24, 
Pasadena, CA 91125, Email: ll@astro.caltech.edu}
\altaffiltext{3}{Hubble Fellow}
\altaffiltext{4}{Department of Astronomy, University of Wisconsin, 475 N. 
Charter Street, Madison, WI 53706}
\altaffiltext{5}{Department of Physics and Astronomy, Johns Hopkins University, 
3400 N. Charles Street, Baltimore, MD 21218}
\altaffiltext{6}{CSIRO, Australia Telescope National Facility, P.O. Box 76, 
Epping, NSW 2121, Australia}

\begin{abstract}

   We estimate the abundances of S and Fe in the high velocity cloud
HVC 287.5+22.5+240, which has a velocity of +240 km s$^{-1}$ with respect
to the local standard of rest and is in the Galactic direction 
$l\sim287^0$ and $b\sim 23^0$.  The measurements are based on UV absorption 
lines of these elements in the Hubble Space Telescope spectrum of NGC 3783, 
a background Seyfert galaxy, as well as new H I 21-cm interferometric
data taken with the Australia Telescope.
We find S/H=$0.25\pm0.07$ and Fe/H=$0.033\pm0.006$ solar, with 
S/Fe=$7.6\pm2.2$ times the solar ratio. The S/H value provides an
accurate measure of the chemical enrichment level in the HVC, while the 
super-solar S/Fe ratio clearly indicates the presence of dust,
which depletes the gas-phase abundance of Fe.
The metallicity and depletion information
obtained here, coupled with the velocity and the position of the HVC
in the sky, strongly suggest that the HVC originated from the Magellanic
Clouds. It is likely (though not necessary) that the same process(es) that
generated the Magellanic Stream is responsible for HVC287.5+22.5+240.

\end{abstract}

\keywords{ galaxies: individual (NGC3783) - galaxies: Seyfert
- Galaxy: halo - ISM: abundances}

\section{INTRODUCTION}

High velocity clouds (HVCs) are H I clouds moving at velocities 
inconsistent with simple models of Galactic rotation and are generally
defined to  have $\vert v_{LSR}\vert$ exceeding 90 km s$^{-1}$ 
(e.g., Wakker \& van Woerden 1997).
HVCs cover a significant fraction of the sky: about 37\% at a H I
column density limit of $7\times 10^{17}$ cm$^{-2}$ (Murphy, Lockman,
\& Savage 1995).
Though HVCs were traditionally found only by their H I emission,
there are now several cases of high velocity gas seen in absorption
in C IV and Si IV (Sembach et al 1995, 1997; Savage, Sembach, \& Lu 1995)
or in Ca II/Mg II (D'Odorico, Pettini, \& Ponz 1985; Meyer \& Roth 1991;
Bowen et al 1994; Ho \& Filippenko 1995) without counterparts in H I emission.

HVCs represent an important Galactic or Local Group
constituent that is poorly understood.
Their origin remains largely unknown even today, decades
after their initial discovery.  The only exception is the Magellanic
Stream, which is generally accepted to be gas pulled out of the
Magellanic Clouds either through tidal interactions with the Milky Way
or by drag forces due to an extended tenuous corona in the Galactic halo.
Many models have been proposed for the origin of HVCs, ranging from disk
gas accelerated to high velocities by energetic events to infalling
primordial gas from extragalactic space (see Wakker \& van Woerden 1997 for
a review). Understanding the HVC
phenomenon may provide valuable information about the interactions
between the Galactic disk and halo and about the evolution of the
Galaxy.

Two of the most important parameters for understanding the origin of HVCs
are their distances and chemical composition.  Recent efforts using 
stellar and extragalactic
probes to detect the optical and UV absorption lines from HVCs have provided
useful constraints on the distances of several HVCs 
(see Wakker \& van Woerden 1997
for a summary and for references), which indicate that several major
HVC complexes are at least 4 kpc distant and at least 2.5 kpc 
above the Galactic plane.
Absorption line studies at the ultraviolet (UV) and optical wavelengths
also indicate that 
most, and perhaps all, HVCs contain some heavy elements (cf. 
Wakker \& van Woerden 1997).
However, deriving accurate elemental abundances for HVCs
has proven difficult for several reasons: (1) one needs
relatively high resolution (FWHM$<20$ km s$^{-1}$) data in order to 
separate the HVC absorption from the absorption at lower velocities 
arising from Milky Way disk gas, as well as to derive accurate
column densities; such data are difficult to obtain
in the UV; (2) optical Ca II and Na I absorption lines, which are much easier
to observe, do not provide accurate abundance estimates
since these ions are not the dominant ionization species in H I gas;
(3) most measurements can be affected by elemental gas-phase depletion
due to dust and in principle
yield only lower limits to the actual abundance of the elements; and
(4) the reference H I column densities must be obtained from high
resolution radio observations because of the presence of small angular
scale structures in HVCs.

   In an earlier Hubble Space Telescope (HST) program (Lu et al 1994), we
obtained a UV spectrum of the bright
Seyfert galaxy NGC 3783, which is projected onto a high velocity cloud
denoted HVC 287.5+22.5+240 at Galactic longitude $l\sim287^o$ and 
latitude $b\sim23^o$ (Mathewson, Cleary, \& Murray 1974; Hulsbosch 1975;
Morras \& Bajaja 1983).
We clearly detected the S II $\lambda\lambda$1250, 1253 absorption lines
from the HVC at $v_{LSR}=+240$ km s$^{-1}$, which allowed us to derive
a sulphur abundance for the HVC. This was the first
HVC abundance measurement free of complications
by dust because S is not readily affected by dust in the Galactic ISM
(Jenkins 1987).
In this paper, we present HST observations of Fe II absorption from
the HVC and use the relative abundances of S and Fe to probe the
dust content of the HVC.  We also present a more reliable measure of
H I column density in the HVC in the direction of NGC 3783 based on
new interferometric radio observations, which is crucial for an accurate
determination of elemental abundances.
The observations and data handling are
presented in section 2. The abundance measurements are described
in section 3. Evidence for dust in the HVC and implications of our
results for the origin of the HVC are discussed in section 4. 
Section 5 briefly summarizes the main conclusions.

\section{OBSERVATIONS AND DATA REDUCTION}

    We obtained a spectrum of NGC 3783 for HST program GO-6500
using the Goddard High Resolution Spectrograph (GHRS)
on 7 July 1996. The G270M grating and the large science aperture 
(2"x2") were used. 
The spectrum was obtained with step-pattern 4, which provides two samples
per diode.  Standard FP-split and comb-addition procedures were used
to reduce fixed pattern noise resulting from irregularities in the
detector window and in the photocathode response. 
The total integration time was 9216 seconds with 11\% of the time
spent on measuring the background.  The spectrum covers the wavelength 
region 2334-2380 \AA, which contains the Fe II $\lambda\lambda$2344 and 
2374 transitions.  
The resulting spectrum has a resolution of FWHM=13 km s$^{-1}$,
which is approximately one diode width.
The signal-to-noise ratio in the continuum is $\sim$14 per diode.
The data are preserved in the HST archive under 
the identification Z38P0101T.

    Data reductions were carried out using the IDL-based GHRS reduction
software (Robinson et al 1992) current as of July 1996. 
A continuum was established
by fitting a cubic spline function to spectral regions free of absorption
lines.  Figure 1 shows the continuum-normalized profiles of
the Fe II $\lambda$2344 and $\lambda$2374 lines recorded in the spectrum,
as well as that of S II $\lambda$1253 obtained by Lu et al (1994).
The conversion between heliocentric velocity and 
LSR velocity is $v_{LSR}=v_{helio}-7.3$ km s$^{-1}$.
 Note that the S II absorption was obtained before the
installation of COSTAR\footnote{COSTAR refers to the corrective optical
system that was installed in the HST during the December 1993 Space Shuttle
service mission.}, and the line spread function had a narrow core
with broad wings (see Lu et al 1994).
We also show in figure 1 the 21-cm emission profile toward NGC 3783
obtained by Murphy et al (1996) with the NRAO 140-foot single-dish
radio telescope with a beam width of 21' at a velocity resolution of
1 km s$^{-1}$. The NRAO data indicate a total H I column 
density of $N$(H I)=$1.15\times 10^{20}$ cm$^{-2}$ for the HVC. 
However, it is known
that HVCs can have complex structures down to 1' scales (cf. Wakker
\& van Woerden 1997). Consequently, this $N$(H I) estimate could differ
considerably from the true line-of-sight $N$(H I) toward NGC 3783.

To determine a more accurate HI column density for the HVC along the
sightline toward NGC3783, we obtained
high angular resolution (1') H I data with the Australia Telescope
Compact Array (ATCA) interferometer at
1 km s$^{-1}$ velocity resolution.
The interferometer data will be discussed in detail in a forthcoming paper
(Wakker et al 1998, in preparation).  We note that the properties of the 
fine structure map of HVC287.5+22.5+240 from the interferometer data
are similar to those observed in other HVCs 
(cf. Wakker \& van Woerden 1997 and
references therein). Variations of a
factor $\sim$5 in H I column density can occur over a few arc minutes, 
highlighting
the danger of estimating line-of-sight $N$(H I) from large-beam
radio observations. About 30\% of the flux in
the NRAO single-disk data is recovered by the interferometer. 
The ATCA data show that the column density directly toward NGC3783
is actually $\sim 8\times 10^{19}$ cm$^{-2}$, as we explain next.

First, the interferometer data were calibrated and mapped in the standard
fashion. Uniform weighting was used in order to obtain the maximum possible
resolution, giving a synthesized
beam of 60"x38". The interferometer is sensitive to structure on scales less
than 30'. Individual velocity channels were 1 km s$^{-1}$ wide. Since the
aperture plane coverage is fairly uniform, the major artifact in the maps
is the presence of a ``bowl'' due to the non-observed zero and short spacings.
To correct for this, each channel map was deconvolved with the synthesized
beam using the Multi-Resolution Clean method (Wakker \& Schwarz 1988).
The channel maps were then corrected for the primary beam of the ATCA, which
has an FWHM of 34'. To compare the results with the NRAO 140-foot spectrum,
a simulated spectrum was generated by ``observing'' the final interferometer
map with the NRAO single-dish (assuming a Gaussian beam of FWHM=21'), and
converting the spatially-integrated flux to brightness temperature using
a conversion factor of 0.38 K/Jy. The simulated spectrum represents the
flux recovered by the interferometer within the 21' beam of the NRAO single
dish.

 Figure~2 shows the spectrum of the HVC from the NRAO single-dish (21' beam) 
and the (simulated) spectrum recovered by the ATCA interferometer at 1' 
resolution. The difference between the two represents the flux filtered
out by the interferometer.
The single-dish spectrum can be adequately modeled as the superposition of
two Gaussian components with a total $N$(H I) of 
$1.15\times 10^{20}$ cm$^{-2}$.
The spectrum recovered by the interferometer shows three components with a
total $N$(H I) of $3.4\times 10^{19}$ cm$^{-2}$.
Thus, emission with an average column density of $8.1\times 10^{19}$ cm$^{-2}$
was filtered out by the interferometer.
Since in the exact direction of NGC 3783 there is essentially
no emission in the interferometer map, the above average column density
also represents approximately the $N$(H I) toward NGC 3783.  
The exact $N$(H I) toward NGC 3783 may differ from the 
above average column density owing to variation on scales greater than 30'. 
However, experiments show that such variations 
should be less than $1.0\times 10^{19}$
cm$^{-2}$. We therefore adopt
$N$(H I)=$(8\pm1) \times 10^{19}$ cm$^{-2}$ or log $N$(H I)=$19.90\pm0.05$
in the analysis.

\section{ABUNDANCE MEASUREMENTS}


     To estimate the column density of Fe II in the HVC, we show in figure 3
the apparent column density profiles, $N_a(v)$, of the 
Fe II $\lambda\lambda$2344, 2374 lines from the HST spectrum, which give the
column density per unit velocity smeared by the line spread function 
(see Savage \& Sembach 1991 for a detailed discussion of $N_a(v)$). 
We adopt the Fe II oscillator strengths from Morton (1991) for the
$\lambda$2344 line ($f=0.11$) and from Cardelli \& Savage (1995) and
Bergeson, Mullman, \& Lawler (1996) for the $\lambda$2374 line ($f=0.0329$).
The $N_a(v)$ profiles of the two Fe II lines (figure 3), 
which differ in strength ($f\lambda$) by a factor 
of 3.3, agree well to within the measurement uncertainties, which suggests that
neither line contains hidden, saturated components that are not resolved.
It is thus straightforward to integrate the $N_a(v)$ profiles to obtain total
column densities. Over the velocity interval 180-300 km s$^{-1}$, we find
log $N$(Fe II)=$13.95\pm0.13$ and $13.93\pm0.05$ from the $\lambda$2374 
and $\lambda$2344 lines, respectively. Using different velocity intervals
does not appreciably change the total column densities. For example,
over 200-280 km s$^{-1}$ 
we find log $N$(FeII)=$13.88\pm0.11$ and $13.89\pm0.04$,
while over 160-320 km s$^{-1}$ we find log $N$(Fe II)=$14.01\pm0.16$
and $13.95\pm0.05$. We will therefore adopt log $N$(Fe II)=$13.93\pm0.05$
or $N$(Fe II)=$(8.5\pm1.0)\times 10^{13}$ cm$^{-2}$ for the HVC.  

The $N$(H I) of the HVC implies that the gas is very optically thick at
the Lyman limit ($\tau_{LL}\sim 500$).
Since Fe II is the dominant ionization stage of Fe in H I gas,
the ratio of $N$(Fe II)/$N$(H I) should provide a reasonable estimate of
the Fe abundance. We thus find Fe/H=$0.033\pm0.006$ solar or
[Fe/H]$=-1.48\pm0.07$ for the HVC, where [Fe/H]
is the logarithmic abundance of Fe relative to the solar abundance given by
[Fe/H]=log(Fe/H)$_{\rm HVC}-$log(Fe/H)$_{\odot}$. The abundance of S in the HVC
is estimated to be  S/H=$0.25\pm0.07$ solar 
or [S/H]=$-0.60^{+0.11}_{-0.15}$.  The S abundance
given here is 1.7 times higher than that of Lu et al (1994) for two reasons.
(1) Lu et al estimated $N$(S II)=$3.4\times 10^{14}$ cm$^{-2}$ for 
the HVC assuming the S II absorption is on the linear part of curve of growth.
Direct integrations of the $N_a(v)$ profiles of the S II lines from 180
to 300 km s$^{-1}$ yield log $N$(S II)=$14.57^{+0.19}_{-0.36}$
and $14.57^{+0.11}_{-0.15}$ from the $\lambda$1250 and $\lambda$1253 
absorption lines, respectively. 
We have adopted the more accurate value from the
$N_a(v)$ integrations in this paper: log $N$(S II)=$14.57^{+0.10}_{-0.14}$
or $N$(S II)=$(3.7\pm1.0)\times 10^{14}$ cm$^{-2}$. (2) The $N$(H I) of
the HVC adopted here is 1.5 times lower than the one used by Lu et al 
($1.21\times 10^{20}$ cm$^{-2}$), which was  based 
on lower-resolution 21-cm data. In the above calculations,
we have adopted the solar abundances of log (S/H)$_{\odot}=-4.73$
and log (Fe/H)$_{\odot}=-4.49$ from Anders \& Grevesse (1989). 

   In the above analyses, we have assumed that ionization
corrections are sufficiently small that the ratios $N$(S II)/$N$(H I)
and $N$(Fe II)/$N$(H I) yield reasonable estimates of the S and Fe
abundances in the HVC. In reality, 
the HVC may contain some or even a substantial 
amount of ionized gas; failure to account for this gas can
result in large errors in the abundance estimates.
Savage \& Sembach (1996) have discussed various ionization
processes that can have significant effects on interstellar abundance 
determinations. However, because HVC 287.5+22.5+240 is likely to be at
large distances away from the Galactic plane and is not known to contain
stars, the processes that normally dominate the ionization of the gas in the
Galactic disk and low halo may be less important in the HVC. Rather,
the dominant ionization process in the HVC is likely to be photoionization
by UV photons from the extragalactic background. For these reasons,
we have run simple photoionization calculations using the code CLOUDY
(v90.02; Ferland 1996) assuming a uniform 
gas density and a plane-parallel geometry
for the cloud with photons incident from one side. We assume
solar relative abundances (Anders \& Grevesse 1989) for the elements   
and an absolute metallicity of [Z/H]$=-0.6$ based on the S abundance. 
The UV ionizing
background is assumed to have a energy distribution similar to the one
adopted by Madau (1992) without taking into account absorption
by intergalactic clouds. We adopt a value of $J_{\nu}=5\times 10^{-23}$
ergs s$^{-1}$ cm$^{-2}$ Hz$^{-1}$ sr$^{-1}$ 
for the intensity of the UV 
background at the Lyman limit, although the exact value is not
critical since it is the ionization parameter, 
$\Gamma$ (=the logarithmic ratio of
the ionizing photon density to the gas particle density), that determines
the ionization structure of the cloud. Limited constraints on the ionization
state of the HVC gas are provided by the following column densities:
log $N$(S II)$\simeq 14.57$, log $N$(Fe II)$\simeq13.93$, 
log $N$(C IV)$\leq 13.24$, and log $N$(N V)$\leq 13.27$, with the latter two
from the 2$\sigma$ upper limits obtained by Lu et al (1994). The 
CLOUDY results indicate that all constraints are met if $\Gamma<-3.5$,
at which S/H$\simeq N$(S II)/$N$(H I) and Fe/H$\simeq N$(Fe II)/$N$(H I)
to within 0.1 dex. Hence,
within the context of this simple, naive photoionization calculation,
it appears that our S and Fe abundances obtained above are not subjected
to significant ionization corrections. Even assuming S/C$\sim$2-3 times
solar (as is observed in Galactic halo stars; see Wheeler, Sneden, \& Truran
1989) in the HVC will not change the above conclusion. 
However, other ionization processes may also be at work and 
may even play a dominant role. For example, the HVC may be 
photoionized by UV photons leaking out the Galactic disk, or the HVC may
be collisionally ionized due to interactions with hot
diffuse halo gas.  Future observations of 
adjacent ion stages of some elements 
(e.g., C II-IV, Si II-IV, S II-III, etc) will
help to better constrain the ionization properties of the HVC.

\section{DISCUSSION}

\subsection{Evidence for Dust in the HVC}

   The abundance ratio of S and Fe in the HVC, 
[S/Fe]=$0.88^{+011}_{-0.15}$ dex, is a factor 
of $7.6\pm2.2$ higher than the solar value. Super-solar S/Fe ratios are
often observed in Galactic ISM clouds because S is essentially unaffected
by dust while Fe gets incorporated into dust grains easily (cf. Jenkins 1987).
For example, in the cool diffuse disk cloud toward $\zeta$ Ophiuchi,
[S/H]$\simeq 0$ and [Fe/H]$\simeq -2.3$ with a S/Fe ratio
200 times that of the solar value (see Table 5 of Savage \& 
Sembach 1996 and references therein), 
implying that 99.5\% of the Fe atoms in the cloud are missing from the
gas phase and are (presumably) incorporated into dust grains. 
In the warm diffuse disk clouds which is more common, Savage \& Sembach
(1996) report a mean S/Fe$\sim$16.  Even in diffuse
halo clouds, a population of interstellar clouds with the lowest known
level of depletion, Fe is depleted by a factor of 4 (i.e., S/Fe=4 times
the solar value; Sembach \& Savage 1996).
The high S/Fe ratio in HVC 287.5+22.5+240 is likely caused by the 
same dust depletion effect.  The inferred depletion level of Fe 
in the HVC will depend on assumptions about the intrinsic 
S/Fe ratio in the HVC. If the HVC has an intrinsically solar S/Fe ratio,
then Fe is depleted by a factor of $\sim8$ (i.e., 7 out of every 8 Fe
atoms are locked up in grains).  
On the other hand, if the HVC has an intrinsic S/Fe ratio that is similar
to metal-poor Galactic halo stars, where S/Fe$\simeq$2.5 solar (cf. Wheeler,
Sneden, \& Truran 1989), then the implied  Fe depletion is only a factor
of $\sim 3$.  In either case, however, some depletion of Fe is
required to explain the observed high S/Fe ratio. 

   Attempts to search for dust in HVCs through its thermal emission in the
infrared have all produced negative results 
(see review by Wakker \& van Woerden
1997). Such searches are severely hampered by the strong contamination from
dust emission associated with foreground gas in the disk
and low halo. Probing the dust content of HVCs through the effects of dust
on elemental abundances appear to be more fruitful, 
as evidenced by this study. However, this type of study
is limited to HVCs with suitable background probes and requires assumptions
about the reference abundances. As discussed
below, HVC 287.5+22.5+240 probably has its origin in the Magellanic Clouds.
The low dust-to-gas ratios in the SMC, 
$N_{HI}/E_{B-V}=10^{23}$ atoms cm$^{-2}$ mag$^{-1}$, and in the LMC,
$N_{HI}/E_{B-V}=2\times 10^{22}$ atoms cm$^{-2}$ mag$^{-1}$, 
(Koornneef 1984), imply a reddening of 
$E_{B-V}\sim$0.0008-0.004 for HVC 287.5+22.5+240. 
Consequently, it will be extremely
difficult to detect the dust in the HVC via its extinction. 

\subsection{Implications for the Origin of the HVC}

   Information on the abundances and dust content in HVCs can provide valuable
clues to their origin. HVC287.5+22.5+240 
is at the tip of a HVC complex which, projected onto the sky,
appears to originate from the Magellanic Clouds but is on the opposite
side of the Magellanic Stream with respect to the Magellanic Clouds (see
figures 2 and 3 of Mathewson et al 1974). Based on this positional 
coincidence, Mathewson et al (1974) 
suggested that HVC 287.5+22.5+240 may be part of the 
Magellanic Stream produced by
tidal interactions between the Magellanic Clouds and the Milky Way. 
West et al (1985) reported a detection of Ca II absorption associated
with HVC287.5+22.5+240 and deduced a low metallicity 
($N$(Ca II)/$N$(H I)$\sim 1/500$ of the solar Ca/H value) 
for the HVC. West et al argued that 
the low metallicity and the high galactocentric velocity of 
HVC 287.5+22.5+240 make it unlikely that the gas 
originated within the Galactic plane. Rather, they suggested that
the HVC is probably an independent extragalactic object, possibly associated
with the Magellanic Stream. The low metallicity of the HVC was
confirmed by the S abundance determination of Lu et al (1994).
The S abundance determination is significant because S 
is not readily affected by dust in the ISM (Savage \& Sembach
1996 and references therein), while more 
than 90\% of the Ca atoms are generally found in dust
grains (Phillips, Pettini, \& Gondhalekar 1984).

The new S and Fe abundance measurements obtained here can shed new light
on the origin of HVC287.5+22.5+240.  The new measurements have
the added significance that they are referenced to a more reliable H I
column density obtained from the high resolution interferometric data.
The low [S/H]$=-0.6$ found for the HVC is broadly 
consistent with the S abundance
found in the Magellanic Clouds: [S/H]$\simeq -0.57$ for the LMC and
[S/H]$\simeq -0.68$ for the SMC
(Russell \& Dopita 1992). The S/Fe ratio
in the HVC is also similar to that found in the interstellar gas of
the Magellanic Clouds.  For example, Welty et al (1997; 
see also Roth \& Blades 1997) presented to date the most 
extensive and accurate observations of {\it interstellar} gas-phase abundances
in a H I region in the SMC, and found [Zn/H]$\simeq-0.7$
and [Zn/Fe]$\simeq 0.6$ (note that Zn, like S, suffers little dust depletion
in diffuse Galactic ISM clouds). Since the intrinsic abundance pattern in the 
Magellanic Clouds obtained from
stellar abundance studies appears similar to the solar abundance pattern to 
within 0.2 dex or so (Russell \& Dopita 1992), one expects [S/H]$\simeq-0.7$
and [S/Fe]$\simeq 0.6$ dex for the interstellar gas 
in the SMC. The similarities in the
absolute S abundance and in the S/Fe ratio (i.e., the level of Fe depletion)
between HVC 287.5+22.5+240 and the Magellanic Clouds gas provide
additional support to the suggestion that the HVC may have been part of 
the LMC/SMC system previously. 

  One of the main criticisms of the tidal model for the Magellanic Stream
(summarized by Moore \& Davis 1994) has always been that tidal interactions
would naturally lead to both a trailing tail and a leading
arm. While the trailing tail was identified with the Magellanic Stream,
no evidence for the leading arm was evident. However,
recent model calculations by Gardiner \& Noguchi (1996) predict that the
leading arm falls in the range $l=270^o$ to $310^o$, $b=-30^o$ to $60^o$,
$v=100$ to $200$ km s$^{-1}$, with most of the arm at $b=40^o$ to $60^o$.
The observed HVC distribution in this part of the sky
does not fit this arm precisely, but there are
many HVCs in the region $l=270^o$ to $310^o$, $b=0^o$ to $30^o$,
and $v=100$ to $250$ km s$^{-1}$, including the clouds defined by Wakker \&
van Woerden (1991) as population EP (containing HVC287.5+22.5+240)
and complex WD. Gardiner \& Noguchi note that the precise position and
velocities of the leading arm depend on the potential of the LMC, which
they assumed to be spherical and constant in time.
In light of our abundance
and depletion information, it may be worthwhile to revisit some of the
tidal models to see if detailed agreement can be made between the leading
arm and the observed HVC distributions in the general region of
of HVC287.5+22.5+240.

  Very recently, Blitz et al (1996) proposed that HVCs are most plausibly
explained as members of Local Group of galaxies, essentially gas left over
from the formation of the Local Group. Evidence cited include: (1) the velocity
centroid of the HVCs (after excluding the Magellanic Stream) is similar to
that of the Local Group; (2) if the clouds are stable entities and are
gravitationally bound to the Local Group, theoretical considerations place
them at distrances consistent with members of the Local Group; (3) the
clouds exhibit a relation betwen their angular sizes and their velocities
relative to the Local Group Standard of Rest, with clouds inferred to be
closer to the Milky Way having larger angular sizes. It was concluded that
no other models can explain all these observed properties.
However, the S abundance obtained
for HVC287.5+22.5+240 appears too high to be consistent with the (almost) 
primordial material as required in the Blitz et al model. Clearly, it will
be extremely important to obtain reliable metallicity determinations for
a sample of representative HVCs in order to test this model rigorously.


\section{SUMMARY}

    We present a  HST GHRS spectrum of the Seyfert galaxy NGC 3783 
in the spectral region 2334-2380 \AA\ in order to determine the abundance of Fe
in the high velocity cloud HVC287.5+22.5+240, which is projected
onto the background Seyfert galaxy. The HVC has a velocity of +240 km s$^{-1}$
with respect to the local standard of rest and is in the Galactic
direction $l\sim 287^o$ and $b\sim 23^o$. To obtain an accurate H I reference
for the HVC, we also obtained a high-resolution (1')
21-cm map of the HVC using the Australia Telescope Compact Array interferometer.
Coupling the Fe abundance obtained
here  with our earlier determination of S abundance in the HVC, 
we discuss evidence
for the presence of dust in the HVC and implications for the origin of the HVC.
Our main conclusions are as follows:

1. Our high-resolution interferometer data indicate that the actual H I
column density in the HVC along the NGC3783 sightline is 
$(8\pm1)\times 10^{19}$
cm$^{-2}$, which is a factor of 1.5 lower than that  obtained
previously from lower resolution (21' beam) data.  
This result highlights one of
the difficulties in abundance studies of HVCs: because HVCs
usually contain very complex fine structures down to 1' scales,
it is difficult to obtain accurate measures of line-of-sight
$N$(H I) toward background probes unless one obtains high angular resolution
radio observations.

2. We find the following metal abundances for the HVC using our
more accurate $N$(H I) determination: 
S/H=$0.25\pm0.07$ and Fe/H=$0.033\pm0.006$ 
solar. The S/H value provides an accurate measure of the metallicity
level in the HVC since S is not readily affected by dust
depletion. The super-solar S/Fe ratio, $7.6\pm2.2$ times solar, indicates
that Fe is depleted by dust in this HVC. This result provides 
strong evidence that at least some HVCs contain dust grains.
The results also demonstrate that elemental abundance studies provide
an effective way of probing the dust content of HVCs.

3. The metallicity level and the amount of Fe depletion found in
the HVC is very similar to those found in the interstellar gas of
the Magellanic Clouds.  These similarities, coupled
with the velocity and the position of the HVC in the sky, strongly
suggest that the HVC originated from the Magellanic Clouds. 
The same process(es) that produced the Magellanic Stream may also
be responsible for HVC287.5+22.5+240 and the HVC complex in that general region
of the sky (i.e., population EP and complex WD of Wakker \& van Woerden 1991).
In particular, HVC287.5+22.5+240 may represent gas in the 
``leading arm'' that was predicted in tidal models for the Magellanic Stream 
(eg. Gardiner \& Noguchi 1996). The metallicity of the HVC appears too
high to be consistent with the model of Blitz et al (1996), who suggested
that HVCs are gas left over from the formation of the Local Group. However,
reliable metal abundance determinations for a sample of representative
HVCs are needed in order to test the Blitz et al model rigorously.

\acknowledgements
  
  The authors thank Ed Murphy for providing an electronic
version of the NRAO 140-foot spectrum and an anonymous referee for helpful
comments. 
  This project was supported by NASA through grant number GO-06500.01-95A
from the Space Telescope Science Institute, which is operated by the 
Association of Universities for Research in Astronomy, Inc., under
NASA contract NAS5-26555. LL also acknowledges a Hubble
Fellowship through grant number HF1062.01-94A. BDS and BPW are appreciative
of support from NASA through grant NAG5-1852. 
KRS acknowledges support from NASA grant GO-06412.01-95A.
The Australia Telescope is funded
by the Commonwealth of Australia for operation as a National Facility
managed by CSIRO.

\begin{figure}
\plotone{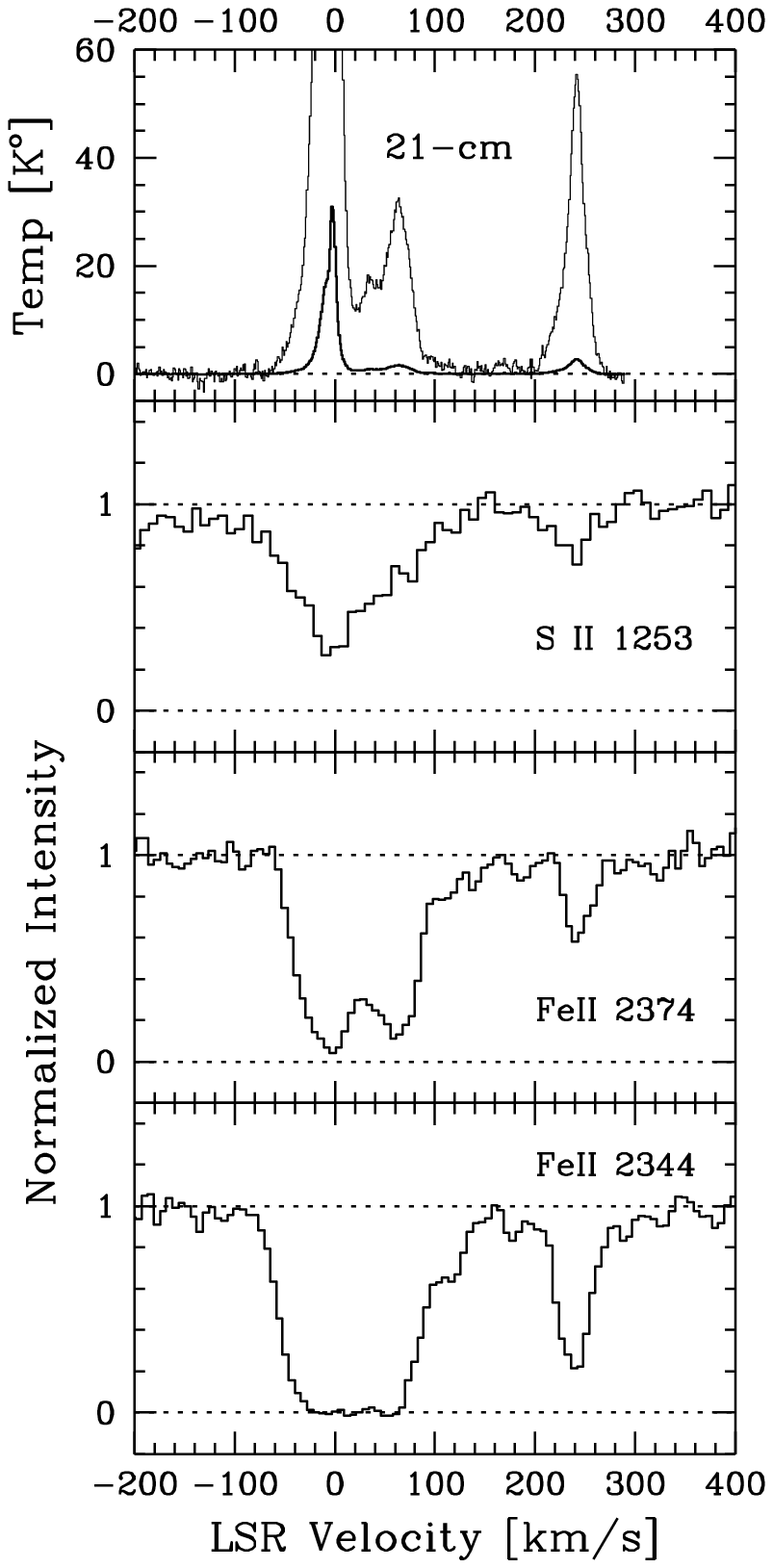}
\caption{The top panel shows the profile of 21-cm 
emission (antenna temperature vs velocity)
toward NGC 3783 taken with the NRAO 140-foot radio
telescope (Murphy et al 1996). The lighter curve is the same profile
expanded by a factor of 20 in the vertical scale.
The remaining panels show the profiles of Galactic ($v<150$ km s$^{-1}$)
and HVC ($v\sim 240$ km s$^{-1}$) absorption toward
NGC 3783 in the lines of S II $\lambda$1253, Fe II $\lambda$2374, and 
Fe II $\lambda$2344. The S II $\lambda$1253 profile was obtained before
COSTAR (see Lu et al 1994) and hence has a somewhat lower resolution than 
that for the Fe II data.}
\end{figure}

\begin{figure}
\plotone{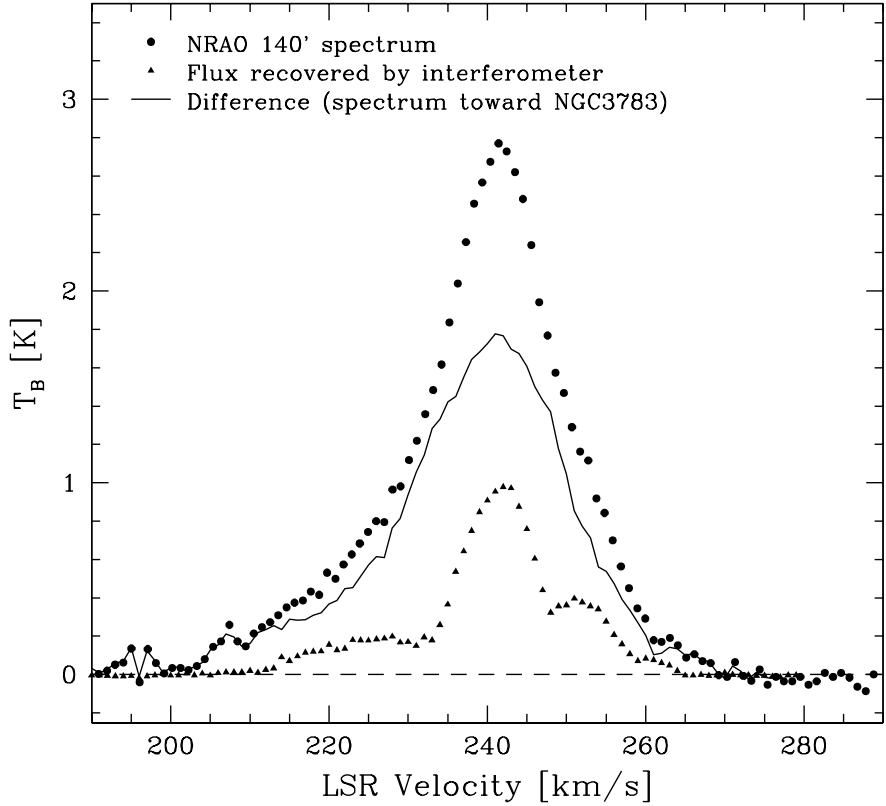}
\caption{Profiles of 21-cm emission from HVC287.5+22.5+240 toward NGC 3783.
Solid dots are observations from the NRAO 140-foot telescope at 21' beam
width, while the solid triangles represent the flux recovered by the 
ATCA interferometer at 1' resolution after convolution with the 21' beam
of the NRAO 140-foot telescope.  The solid curve shows the difference 
between the NRAO observation and the convolved interferometry measurement.
Since no H I emission is seen directly toward NGC3783 in the high
resolution ATCA map, the solid curve represents the appropriate H I
emission to compare to the UV absorption line observations.}
\end{figure}

\begin{figure}
\plotone{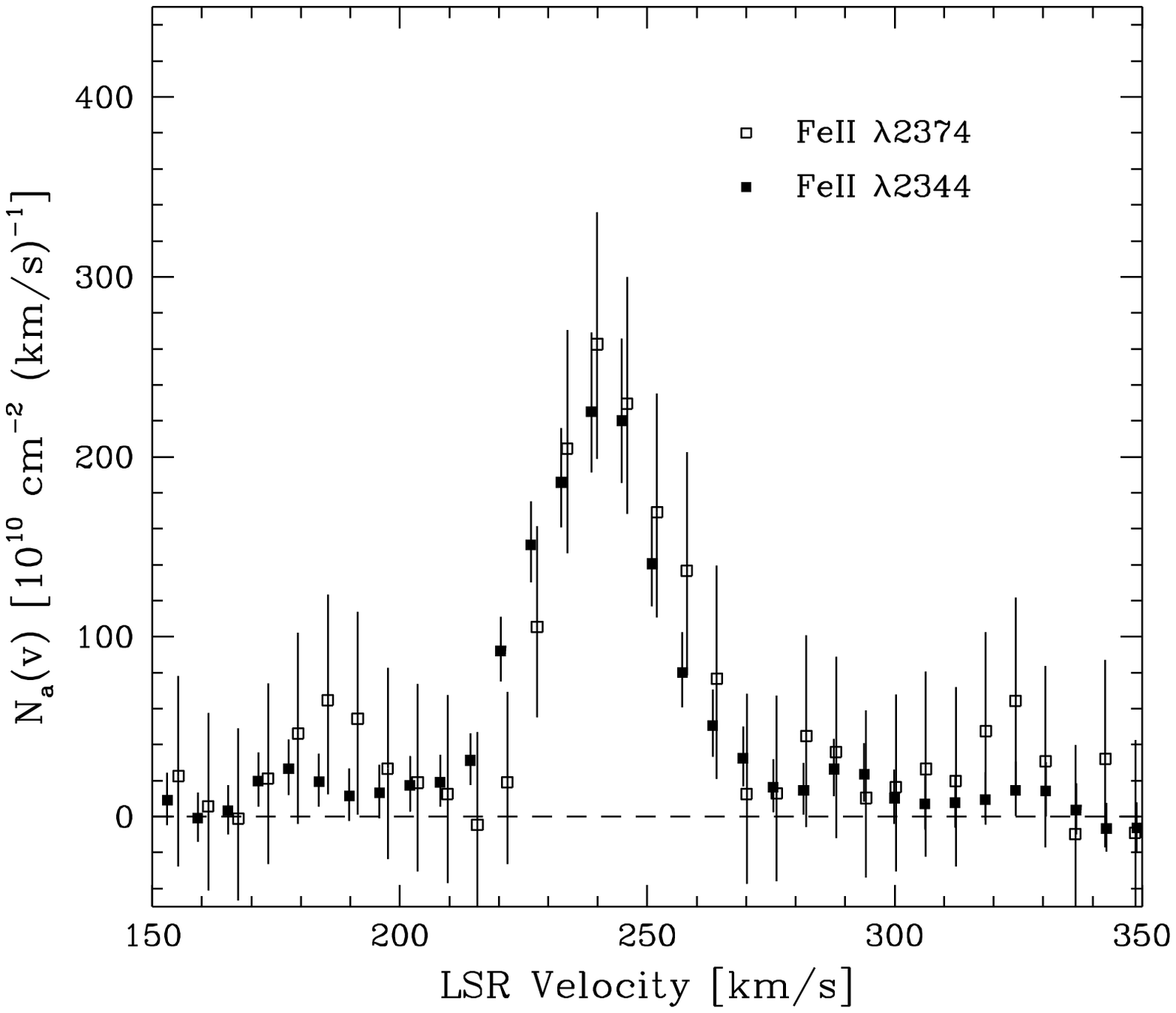}
\caption{Apparent column density profiles of Fe II $\lambda\lambda$2374,
2344 near the HVC absorption at $v_{LSR}=+240$ km s$^{-1}$.}
\end{figure}


\begin{references}

\reference{}
Anders, E., \& Grevesse, N. 1989, in Geochim. Cosmochim. Acta, 53, 197

\reference{}
Bergeson, S.D., Mullman, K.L., \& Lawler, J.E. 1996, ApJ, 464, 1050

\reference{}
Blitz, L., Spergel, D., Teuben, P., Hartmann, D., \& Burton, W.B. 1996,
   BAAS, 28, 1349

\reference{}
Bowen, D.V., Roth, K.C., Blades, J.C., \& Meyer, D.M. 1994, ApJ, 420, L71

\reference{}
Cardelli, J.A., \& Savage, B.D. 1995, ApJ, 452, 275

\reference{}
D'Odorico, S., Pettini, M., \& Ponz, D. 1985, ApJ, 299, 852

\reference{}
Ferland, G. J. 1996, Hazy, a brief introduction to cloudy 90, 
 University of Kentucky,  Physics Department Internal Report

\reference{}
Gardiner, L.T.,  \& Noguchi, M. 1996, MNRAS, 278, 191

\reference{}
Ho, L.C., \& Filippenko, A.V. 1995, ApJ, 444, 165 (erratum in ApJ, 463, 818)

\reference{}
Hulsbosch, A.N.M. 1975, A\&A, 40, 1

\reference{}
Jenkins, E.B. 1987, in Interstellar Processes, Ed. D.J.Hollenbach \&
   H.A.Thronson, Jr. (Dordrecht:Reidel), 533

\reference{}
Koornneef, J. 1984, in Structure and Evolution of the Magellanic Clouds,
 eds. S. van den Berg \& K.S. de Boer (Kluwer), p133

\reference{}
Lu, L., Savage, B.D., \& Sembach, K.R. 1994, ApJ, 426, 563

\reference{}
Madau, P. 1992, ApJ, 389, L1

\reference{}
Mathewson, D.S., Cleary, M.N., \& Murray, J.D. 1974, ApJ, 190, 291

\reference{}
Meyer, D.M., \& Roth, K.C. 1991, ApJ, 383, L41

\reference{}
Moore, B., \& Davis, M. 1994, MNRAS, 270, 209

\reference{}
Morras, R., \& Bajaja,  E. 1983, A\&AS, 51, 131

\reference{}
Morton, D.C. 1991, ApJS, 77, 119

\reference{}
Murphy, E.M., Lockman, F.J., Laor, A., \& Elvis, M. 1996
   ApJS, 105, 369

\reference{}
Murphy, E.M., Lockman, F.J., \& Savage, B.D. 1995, ApJ, 447, 642 

\reference{}
Phillips, A.P, Pettini, M., \& Gondehalekar, P. 1984,
   MNRAS, 206, 337

\reference{}
Robinson, R.D., Blackwell, J., Feggans, K., Lindler, D., 
 Norman, D., \& Shore, S.N.  1992,  
  A User's Guide to the GHRS Software, version 2.0

\reference{}
Roth, K.C., \& Blades, J.C. 1997, ApJ, 474, 95

\reference{}
Russell, S.C., \& Dopita, M.A. 1992, ApJ, 384, 508

\reference{}
Savage, B.D., \& Sembach, K.R. 1991, ApJ, 379, 245

\reference{}
Savage, B.D., \& Sembach, K.R. 1996, ARAA, 34, 279

\reference{}
Savage, B.D., Sembach, K.R., \& Lu, L. 1995, ApJ, 449, 145

\reference{}
Sembach, K.R., \& Savage, B.D. 1996, ApJ, 457, 211

\reference{}
Sembach, K.R., Savage, B.D., Lu, L., \& Murphy, E.M. 1995, ApJ, 451, 616

\reference{}
Sembach, K.R., Savage, B.D., Lu, L., \& Murphy, E.M. 1997, in preparation

\reference{}
Wakker, B.P., \& Schwarz, U.J. 1988, A\&A, 200, 312

\reference{}
Wakker, B.P., \& van Woerden, H. 1991, A\&A, 250, 509

\reference{}
Wakker, B.P., \& van Woerden, H. 1997, ARA\&A, 35, 217

\reference{}
Welty, D.E., Lauroesch, J.T., Blades, J.C., Hobbs, L.M., \& York, D.G. 1997, 
   preprint

\reference{}
West, K.A., Pettini, M., Penston, M.V., Blades, J.C., \& Morton, D.C. 1985,
   MNRAS, 215, 481

\reference{}
Wheeler, J.C., Sneden, C., \& Truran, J.W. 1989, ARA\&A, 27, 279

\end{references}
\end{document}